%% file: main.tex
\documentclass[sigplan,screen]{acmart}
\usepackage{natbib}
\usepackage{booktabs}
\usepackage[acronym,toc,nopostdot,nonumberlist,xindy]{glossaries}
\usepackage{subcaption}
\usepackage{threeparttable}
\usepackage{orcidlink}
\usepackage{svg}
\usepackage[T1]{fontenc}
\usepackage{microtype}
\usepackage{listings}
\usepackage{xcolor}

\lstdefinelanguage{yaml}{
  keywords={true,false,null,y,n},
  keywordstyle=\color{blue},
  sensitive=false,
  comment=[l]{\#},
  commentstyle=\color{gray}\ttfamily,
  stringstyle=\color{purple}\ttfamily,
  morestring=[b]',
  morestring=[b]"
}

\lstdefinestyle{myyaml}{
  language=yaml,
  basicstyle=\scriptsize\ttfamily, 
  backgroundcolor=\color{gray!10},   
  breaklines=true,                  
  frame=single,                     
  frameround=ffff,                  
  framexleftmargin=4pt,             
  showstringspaces=false
}

\newcommand{\yamlinline}[1]{\colorbox{gray!10}{\lstinline[language=yaml,basicstyle=\ttfamily]|#1|}}

\input{glossary}

\input{todos}

\AtBeginDocument{%
  }

\setcopyright{acmlicensed}
\copyrightyear{2026}
\acmYear{2026}
\acmDOI{10.1145/3837730.3837733}
\acmConference[CUG '26]{Cray User Group}{April 26-30, 2026}{Nice, France}

\begin{document}

\title{Multi-tenant Kubernetes Use Cases for AI, Secure Computing and Data Services, and More}


\author{Jake Watson}
\affiliation{
\institution{Bristol Centre for Supercomputing\\
University of Bristol}
\city{Bristol}
\country{United Kingdom}}
\email{jake.watson@bristol.ac.uk}

\author{Sadaf R. Alam\orcidlink{0000-0002-2534-5078}}
\affiliation{
\institution{Bristol Centre for Supercomputing\\
University of Bristol}
\city{Bristol}
\country{United Kingdom}}
\email{sadaf.alam@bristol.ac.uk}
\orcid{0000-0002-2534-5078}

\author{Christopher Woods\orcidlink{0000-0001-6563-9903}}
\affiliation{
\institution{Bristol Centre for Supercomputing\\
University of Bristol}
\city{Bristol}
\country{United Kingdom}}
\email{christopher.woods@bristol.ac.uk}
\orcid{0000-0001-6563-9903}

\author{Abdelwahab Kawafi\orcidlink{0000-0002-1369-6698}}
\affiliation{
\institution{Bristol Centre for Supercomputing\\
University of Bristol}
\city{Bristol}
\country{United Kingdom}}
\email{a.kawafi@bristol.ac.uk}
\orcid{0000-0002-1369-6698}

\author{Thomas Green\orcidlink{0009-0001-8622-3664}}
\affiliation{
\institution{Bristol Centre for Supercomputing\\
University of Bristol}
\city{Bristol}
\country{United Kingdom}}
\email{thomas.green@bristol.ac.uk}
\orcid{0009-0001-8622-3664}

\author{Ian Johnson}
\affiliation{\institution{HPE HPC/AI EMEA Research Lab \\
Hewlett Packard Enterprise}
\city{Bristol}
\country{United Kingdom}}
\email{ian.johnson@hpe.com}

\author{Ellis Pires\orcidlink{0000-0002-6531-8269}}
\affiliation{\institution{HPE HPC/AI EMEA Research Lab \\
Hewlett Packard Enterprise}
\city{Bristol}
\country{United Kingdom}}
\email{ellis.pires@hpe.com}
\orcid{0000-0002-6531-8269}

\author{Jessica R. Jones\orcidlink{0009-0003-7579-7245}}
\affiliation{
\institution{HPE HPC/AI EMEA Research Lab \\
Hewlett Packard Enterprise}
\city{Bristol}
\country{United Kingdom}}
\email{j.r.jones@hpe.com}
\orcid{0009-0003-7579-7245}

\author{Utz-Uwe Haus\orcidlink{0000-0001-7292-9984}}
\affiliation{\institution{HPE HPC/AI EMEA Research Lab \\
Hewlett Packard Enterprise}
\city{Z\"{u}rich}
\country{Switzerland}}
\email{utz-uwe.haus@hpe.com}
\orcid{0000-0001-7292-9984}

\renewcommand{\shortauthors}{Bristol Centre for Supercomputing (BriCS) and HPE}

\begin{abstract}
Kubernetes, as a container orchestration engine, has been widely used in
cloud-native ecosystems for several years. In supercomputing ecosystems,
especially where bare-metal performance for compute and network devices are
considered, the adoption is somewhat limited. However, with the increasing
diversity of use cases such as AI, secure and confidential computing for
sensitive data, and mixed workload orchestration, a traditional, single-tenant
batch computing system does not offer the flexibility and reproducibility to which public cloud users are accustomed. Note that Kubernetes is not considered a replacement for batch scheduling systems, which have powerful features for
large-scale MPI jobs with thousands of network end points. Rather, it is a
complementary service provided as part of a national \acrlong{AIRR}. We evaluate
Kubernetes deployment on a \gls{HPE} Cray EX supercomputer with HPE Slingshot
interconnect, called Isambard-AI, with co-design use cases. One is a \acrlong{TRE} used
for medical and health sciences. The other combines KubeRay, Ray, and vLLM to
provide a distributed, sandboxed, persistent AI model hosting service targeting
multi-tenant confidential computing. We discuss challenges and lessons learned,
and where further development is needed to offer a production
Kubernetes-as-a-Service on \gls{HPE} Cray EX (and later) platforms.
\end{abstract}



\begin{teaserfigure}
  \vspace{13mm}
  \Description{}
\end{teaserfigure}

\maketitle

\section{Introduction}\label{sec:introduction}
\input{intro-k8s}

\section{Background}\label{sec:background}
\input{bg-k8s}
\input{bg-FRIDGE}
\input{bg-model-hosting}

\section{Implementation}\label{sec:implementation}
\input{impl-versions-baseline}
\input{impl-VNI-isolation}
\input{impl-FRIDGE}
\input{impl-model-hosting}

\section{Evaluation and Results}\label{sec:evaluation}
\input{eval-K8s}

\input{eval-FRIDGE}
\input{eval-model-hosting}

\section{Summary and future work} \label{sec:summary}
\input{summ-conc} \label{conc}
\input{summ-future} \label{future}

\section*{Acknowledgments}

We would like to thank teams at the \gls{BriCS} and \gls{HPE} for their contributions and support that involves not only setting up the infrastructure for the Kubernetes services but also the wider AI software stack. The \gls{FRIDGE} project partners at the Alan Turing institute, University College London, and University of Cambridge have been instrumental in co-designing the architecture on the UK \gls{AIRR} resources including Isambard-AI and public cloud ecosystems. Isambard-AI is funded by the UK Government’s \gls{DSIT} via \gls{UKRI} and \gls{STFC}. The \gls{FRIDGE} project has been funded by the \gls{DARE UK} Early Adopters programme.  

\typeout{}

\bibliography{brics-bib}
\printglossaries

\end{document}

%% file: glossary.tex
\setacronymstyle{long-short}
\makeglossaries

\newacronym{AI}{AI}{Artificial Intelligence}
\newacronym{AIRR}{AIRR}{AI Research Resource}
\newacronym{AISI}{AISI}{AI Security Institute}
\newacronym{API}{API}{Application Programming Interface}
\newacronym{ARP}{ARP}{Address Resolution Protocol}
\newacronym{BGP}{BGP}{Border Gateway Protocol}
\newacronym{BriCS}{BriCS}{Bristol Centre for Supercomputing}
\newacronym{CFS}{CFS}{Cray Framework Service}
\newacronym{CNCF}{CNCF}{Cloud Native Computing Foundation}
\newacronym{CNI}{CNI}{Container Network Interface}
\newacronym{CPU}{CPU}{Central Processing Unit}
\newacronym{CRD}{CRD}{Custom Resource Definition}
\newacronym{CSM}{CSM}{Cray System Management}
\newacronym{CXI}{CXI}{Cray eXascale Interconnect}
\newacronym{DARE UK}{DARE UK}{Data and Analytics Research Environments UK}
\newacronym{DRA}{DRA}{Dynamic Resource Allocation}
\newacronym{DRC2}{DRC2}{Dynamic Resource Control 2}
\newacronym{DSIT}{DSIT}{Department of Science, Innovation and Technology}
\newacronym{FRIDGE}{FRIDGE}{Federated Research Infrastructure by Data Governance Extension}
\newacronym{GPU}{GPU}{Graphical Processing Unit}
\newacronym{HA}{HA}{High Availability}
\newacronym{HPC}{HPC}{High-Performance Computing}
\newacronym{HPE}{HPE}{Hewlett Packard Enterprise}
\newacronym{HSN}{HSN}{High-Speed Network}
\newacronym{I-AI}{IS-AI}{Isambard-AI}
\newacronym{ITL}{ITL}{Inter Token Latency}
\newacronym{k8s}{K8s}{Kubernetes}
\newacronym{KV}{KV}{Key-Value}
\newacronym{LLM}{LLM}{Large Language Model}
\newacronym{MAP}{MAP}{Mutating Admission Policy}
\newacronym{ML}{ML}{Machine Learning}
\newacronym{MPI}{MPI}{Message Passing Interface}
\newacronym{MPMD}{MPMD}{Multiple Program, Multiple Data}
\newacronym{MTBF}{MTBF}{Mean Time Between Failures}
\newacronym{NCCL}{NCCL}{NVIDIA Collective Communications Library}
\newacronym{NIC}{NIC}{Network Interface Card}
\newacronym{OFI}{OFI}{Open Fabrics Interfaces}
\newacronym{RBAC}{RBAC}{Role-Based Access Control}
\newacronym{RDMA}{RDMA}{Remote Direct Memory Access}
\newacronym{RKE2}{RKE2}{Rancher Kubernetes Engine 2}
\newacronym{RPC}{RPC}{Remote Procedure Call}
\newacronym{SATRE}{SATRE}{Standard Architecture for Trusted Research Environments}
\newacronym{SDE}{SDE}{Sensitive Data Environment}
\newacronym{SHS}{SHS}{Slingshot Host Software}
\newacronym{STFC}{STFC}{Science and Technology Facilities Council}
\newacronym{TPOT}{TPOT}{Time per Output Token}
\newacronym{TPS}{TPS}{Tokens per Second}
\newacronym{TRE}{TRE}{Trusted Research Environment}
\newacronym{TTFT}{TTFT}{Time to First Token}
\newacronym{UKRI}{UKRI}{UK Research and Innovation}
\newacronym{USS}{USS}{User Services Software}
\newacronym{VNI}{VNI}{Virtual Network Identifier}
\newacronym{VPU}{VPU}{Vector Processing Unit}
\newacronym{WLM}{WLM}{Workload Manager}


%% file: todos.tex
\usepackage[colorinlistoftodos,textwidth=3.7cm,textsize=tiny]{todonotes}

\newcounter{todocounter}

\newcommand{\jrj}[2][]%
{\stepcounter{todocounter}\todo[color=green!40,#1]{\thetodocounter: JRJ : #2}}

\usepackage[most]{tcolorbox}

%% file: intro-k8s.tex
Traditionally, \gls{HPC} systems have been designed for running tightly-coupled, large-scale batch workloads. There is now an increasing demand from end-users that \gls{HPC} systems support increasingly diverse workloads including AI model training and inference, secure AI system sandboxing, and analysis of sensitive data with strict governance requirements. Batch schedulers like Slurm are incredibly powerful for bulk workloads but they often lack the flexibility, service abstractions, and multi-tenant isolation that is needed to address these advanced use cases. The integration of cloud-native technologies into \gls{HPC} environments presents a solution to address these use cases but carries a new set of challenges.

Kubernetes is the pre-eminent technology for multi-tenant workload orchestration in cloud environments. The Kubernetes ecosystem is mature and feature-rich, offering tools for autoscaling, load balancing, namespace isolation, identity federation, software-defined networking, and infrastructure-as-code. The integration of Kubernetes onto bare-metal \gls{HPC} systems, particularly those with advanced high-speed fabrics, is non-trivial and requires modifications that are not well-supported in the community. We document the design, deployment, and evaluation of Kubernetes-as-a-Service on \gls{I-AI}, the UK's national sovereign \gls{AIRR}, which comprises an \gls{HPE} Cray EX framework with the Slingshot interconnect and NVIDIA Grace (ARM64) Hopper superchips. The Kubernetes-based services presented here on \gls{I-AI} are intended to complement, not replace, the Slurm batch scheduler; Slurm is still expected to schedule the largest tenants and partitions, while Kubernetes services target niche and ever-evolving AI and sensitive data handling use cases.

The increasing demand of Slurm and Kubernetes represents a growing effort to bridge traditional \gls{HPC} environments with modern cloud-native workloads, though significant technical and organisational challenges remain. Projects like Slinky~\cite{slinky}, developed by SchedMD and NVIDIA, aim to natively integrate Kubernetes with Slurm, allowing containerized workloads to be scheduled alongside traditional batch jobs; this approach introduces complexity in maintaining consistent resource accounting across two distinct scheduling domains. Similarly, the Slurm Operator project~\cite{kubeoperator} provides a Kubernetes operator that can submit and manage Slurm jobs from within a Kubernetes cluster, yet it struggles with latency overhead and the impedance mismatch between Kubernetes' declarative, eventually-consistent model, and Slurm's tightly controlled, synchronous job lifecycle. The \gls{CNCF} Batch System Initiative Working Group~\cite{cncfbatch} has worked to standardise batch scheduling interfaces, but achieving consensus across diverse \gls{HPC} and cloud-native stakeholders has proven slow and difficult. Tools like Volcano~\cite{volcano} and Armada~\cite{armada} extend Kubernetes with HPC-style scheduling primitives, such as gang scheduling and queue management, but often lack the maturity and production-hardening that Slurm has accumulated over decades. A deeper challenge across all these approaches is the fundamental tension between Kubernetes' dynamic, ephemeral resource model; and Slurm's assumption of stable, dedicated hardware; making seamless interoperability difficult to achieve in practice, particularly for MPI-based workloads, high-speed interconnects like SlingShot, and bare-metal GPU access upon which AI, ML, and HPC applications critically depend.

We therefore implement the multi-tenancy Slingshot network isolation solution from Friese \emph{et al.}~\cite{friese2025closinghpccloudconvergencegap} on \gls{I-AI}~\cite{mcintosh2024isambard}. We build upon this adaptation to deploy Kubernetes-as-a-Service to address two use cases. First, \gls{FRIDGE}~\cite{watson2025code}: a Kubernetes-based, \gls{SATRE}-compliant~\cite{satre} `satellite' \gls{TRE} for researchers analysing data with strict governance requirements. Subsequently, a combination of KubeRay, Ray~\cite{moritz2018ray}, and vLLM~\cite{kwon2023} is used to serve a distributed, sandboxed, persistent, AI model with dynamic resource allocation to accommodate the UK \gls{AISI} user base and the emerging requirement to isolate model execution environments for both security and reproducibility concerns. We evaluate the suitability of this deployment for multi-tenant computing at scale.

The implementation and evaluation of the use cases highlight several areas of improvement that we have identified for further investigation and development.  These include reducing the operational complexity by automating and integrating independent components that are required for a multi-tenant Kubernetes implementation. A sustainable solution may require a CXI Kubernetes \gls{DRA} driver~\cite{dra} that could be maintained as a community. In short, the work presented in this paper provides a solid baseline for hardening the solution for additional Kubernetes use cases for operational environments on existing EX and future GX infrastructures, based on the Slingshot interconnect, which are expected to maintain compatibility with open standards.

The remainder of this paper is structured as follows. Section \ref{sec:background} provides background on the three areas of this work; bare-metal Kubernetes and multi-tenancy, \glspl{TRE}, and secure and multi-node model hosting. Section \ref{sec:implementation} details the implementation of each of these and the specific adaptations required for \gls{I-AI}. Section \ref{sec:evaluation} presents results and evaluations for each area, including benchmarking, assessment of the \gls{TRE} deployment, and analysis of secure model hosting configurations. Finally, Section \ref{sec:summary} summarises the contributions of this work and discusses directions for ongoing and future research and development.

%% file: bg-k8s.tex
\subsection{Bare-metal Kubernetes and Multi-tenancy}

A challenge with bare-metal \gls{k8s} deployments is that they require manual integrations for storage, networking, and load balancing; unlike cloud distributions which benefit from provider-managed integrations. Distributions like K3s and \gls{RKE2}~\cite{rke2} address this by offering a `near-out-of-the-box' experience. However, HPC environments still require site-specific adaptations including: support for immutable operating systems, provisioning load balancer IPs, and integrating high-speed interconnects like Slingshot. We outline the adaptations required to deploy \gls{FRIDGE} efficiently and securely in \gls{I-AI}, using a portable, multi-platform architecture. 

\gls{HPE} Slingshot is a high-speed network fabric that fundamentally comprises Cassini ASIC NICs and Rosetta switches. Ethernet communication is supported through \emph{netdev} Linux devices and \gls{RDMA} is supported via character devices. The \gls{RDMA} plane bypasses the kernel network and does not adhere to standard OS-level network isolation mechanisms. Instead, it uses a separate access control model that relies on the \gls{VNI} domains and the \gls{CXI} services.

\glspl{VNI} are integer labels that define isolated communication domains on the fabric, analogous to VLANs. When enforcement is enabled, Rosetta switches only route packets tagged with a given \gls{VNI} to a destination if it belongs to the same \gls{VNI} domain. The \gls{CXI} services are configured on each node by a privileged service, such as the Slurm \gls{WLM} plugin, and define which principals are allowed to use the \gls{RDMA} interface and which \gls{VNI}s they are allowed to operate on. When an application attempts to open an \gls{RDMA} endpoint through \emph{libfabric}, for example, the \gls{CXI} provider must produce an authorisation key comprising an (\emph{svc\_id}, \emph{vni}) tuple that corresponds to a \gls{CXI} service and the key bearer must be an authorised member of that service. If either of these conditions are not met, then endpoint creation for that service fails and the provider initiates a series of fallbacks to identify a service it is permitted to use. When endpoint creation is successful for a given service, the NIC tags all outgoing packets with the \gls{VNI} for that service. The Rosetta switch enforces whether the traffic can be directed toward the receiving NIC. The receiving NIC will only deliver the packets to the receiving endpoint if it matches the addressing context and \gls{VNI}.

Kubernetes network isolation operates on the IP layer through plugins and network policies, which have no knowledge of the \gls{RDMA} plane. Without adaptations, containers on the same node share access to the host \gls{CXI} devices and do not get assigned distinct \gls{CXI} services, precluding fabric isolation between tenants. Friese \emph{et al.}~\cite{friese2025closinghpccloudconvergencegap} address this by modifying the \gls{CXI} driver, the \gls{CXI} library, and \emph{libfabric} to scope \gls{CXI} services to Linux network namespaces on top of existing user and group scopes. Unlike user and group IDs, network namespace IDs are governed from outside of the containers and cannot be modified from within. This provides a tamper-resistant \gls{CXI} service binding and allows each container its own \gls{CXI} service and \gls{VNI} assignment, enabling container-node multi-tenancy.

%% file: bg-FRIDGE.tex
\subsection{Trusted Research Environments}

Researchers working with sensitive data, such as medical records, require computing environments that enforce strict data governance; access and data egress must be restricted and auditable. \glspl{TRE}, sometimes referred to as \glspl{SDE}, have risen to prominence in recent years to provide these guarantees. However, current \gls{TRE} deployments are often limited to cloud-based infrastructure and carry associated drawbacks such as compliance, cost and procurement, and vendor lock-in. A local \gls{TRE} service on \gls{I-AI} would mitigate many of these issues and provide researchers with accessible and powerful confidential AI computing.

\begin{figure}[htbp]
    \Description{An illustration of the FRIDGE architecture}
    \centering
    \includegraphics[width=\columnwidth]{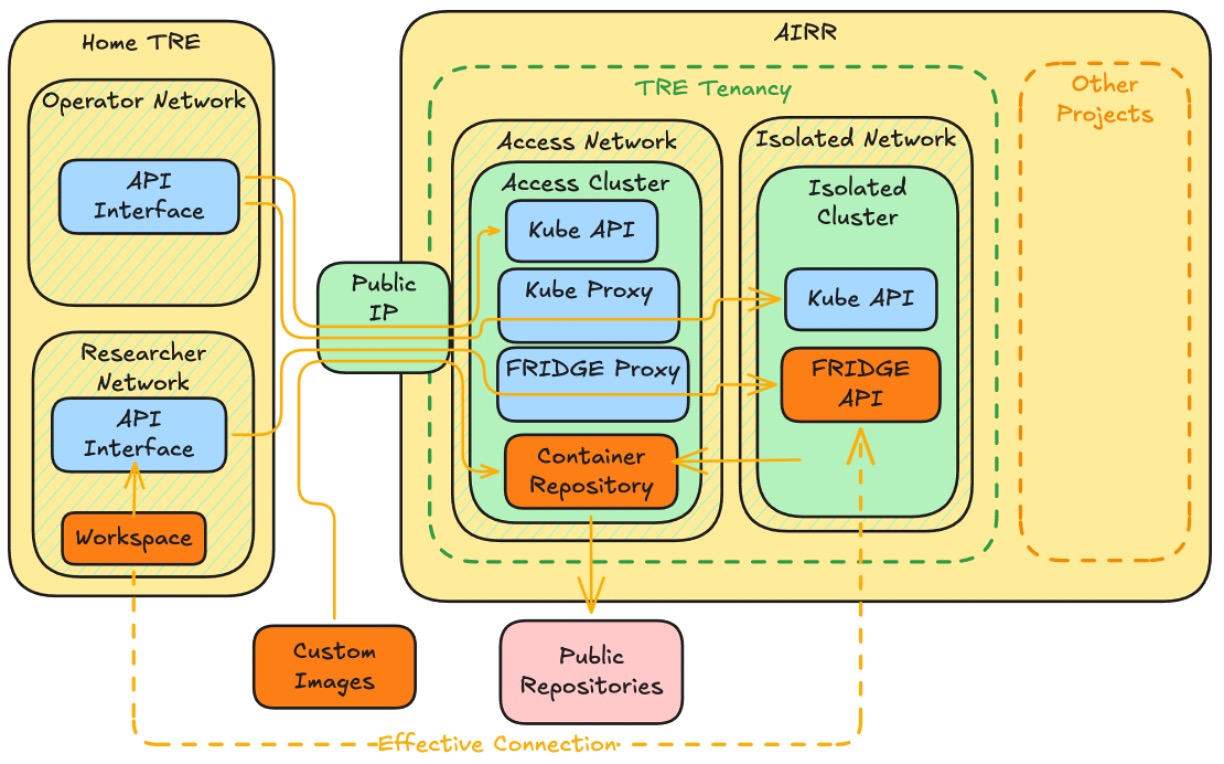}
    \caption{The \gls{TRE} architecture developed by the \gls{FRIDGE} project team from The Alan Turing Institute, University College London, University of Bristol, and University of Cambridge for public-cloud, on-premises and \gls{AIRR} supercomputers including Isambard-AI at Bristol and Dawn at Cambridge. Figure reproduced from Alan Turing Institute~\cite{fridgearch}.}
    \label{fig:fridgearch}
\end{figure}


\gls{SATRE}~\cite{satre} is a compliance framework for \glspl{TRE} developed by \gls{DARE UK}~\cite{dare}; it defines a set of architectural principles, requirements, and recommendations for modern \glspl{TRE}. It provides an objective and robust baseline to measure \gls{TRE} deployments and is rapidly becoming the industry standard.

\gls{FRIDGE}~\cite{fridge} is a \gls{DARE UK} project that seeks to provide a \gls{SATRE}-compliant, Kubernetes-based \gls{TRE} service across \gls{AIRR}'s production compute infrastructure. The project adopts a container-based tenancy model for maximum portability across different systems. It also offers a satellite model wherein a host \gls{TRE} can dispatch workloads into an isolated compute environment and securely retrieve the processed results. Deploying \gls{FRIDGE} on \gls{I-AI} would expose its massive-scale GPU compute resources to sensitive data research, while \gls{I-AI}'s unique ARM64 hardware profile serves to validate the portability and reusability objectives of the \gls{FRIDGE} project itself.

The \gls{FRIDGE} architecture diagram in Figure \ref{fig:fridgearch} illustrates how an existing Home \gls{TRE} extends its governance boundary to an external \gls{AIRR} system, effectively borrowing compute resources while maintaining the \glspl{TRE} security controls. The Home \gls{TRE} contains two internal networks. The Operator Network hosts an API through which \gls{TRE} Operators manage and configure the \gls{FRIDGE} instance. The Researcher Network hosts a dedicated API and a `Workspace' used by Job Submitters, a privileged set of Safe Researchers, to create and dispatch computational jobs. Both networks share a single public IP entry point for all outbound communication with external systems.

The \gls{AIRR} system is managed by the \gls{FRIDGE} Hosting Organisation and its Hosting Provider Administrators. Within \gls{AIRR}, a \gls{TRE} Tenancy carves out a secure enclave that is opaque to both the host system and any other tenancies. This tenancy is divided into two sub-networks. The Access Network contains the Access Cluster, which exposes a Kubernetes API, a Kube Proxy for routing, a \gls{FRIDGE} Proxy that mediates communication with the Isolated Network, and a Container Repository that stores approved container images. The Isolated Network contains the Isolated Cluster, where actual computation takes place, which exposes its own Kube \gls{API} and the core \gls{FRIDGE} API — the heart of the satellite \gls{TRE}. The \gls{FRIDGE} API receives requests forwarded from the \gls{FRIDGE} Proxy and orchestrates workloads within the Isolated Cluster. Notably, the Isolated Cluster has no direct public-facing connection, reinforcing the layered security model.

External to the \gls{AIRR} tenancy, two additional components feed into the system: Custom Images, supplied by the TRE Operator Organisation, are pushed into the Container Repository, while Public Repositories supply base container images. An Effective Connection represents the logical end-to-end path from the Home \gls{TRE}'s Workspace through to the \gls{FRIDGE} API in the Isolated Cluster, abstracting over the network hops and proxy layers in between. The diagram also shows Other Projects within \gls{AIRR}, indicating that the \gls{AIRR} system may host multiple independent tenancies simultaneously, each isolated from one another.

Overall, Figure \ref{fig:fridgearch} captures how \gls{FRIDGE} achieves governance boundary extension: the \gls{TRE} Operator Organisation retains control over data, workloads, and user access, while the \gls{FRIDGE} Hosting Organisation provides the underlying compute infrastructure under a shared responsibility model, without need for data sharing agreements.

%% file: bg-model-hosting.tex
\subsection{Secure and Multi-node Model Hosting}

Driven by demand from the UK's AI researchers, particularly \gls{AISI} user base, we aim to support persistent AI model hosting where Kubernetes orchestrates shared-access inference services against common models, each with a single endpoint. The advantages of persistent model hosting include strong sandboxing, standardisation and reproducibility, and efficient cluster resource sharing. We demonstrate the feasibility of this hosting pattern on \gls{I-AI} and Kubernetes.

\begin{figure}[htbp]
    \Description{An illustration of the architectural layers required for distributed AI model training and workloads; compute substrate, container orchestration, distributed compute engine, training and inference frameworks, and AI workload. The technologies highlighted for use in this paper are Kubernetes, Ray, vLLM and PyTorch.}
    \centering
    \includegraphics[width=\columnwidth]{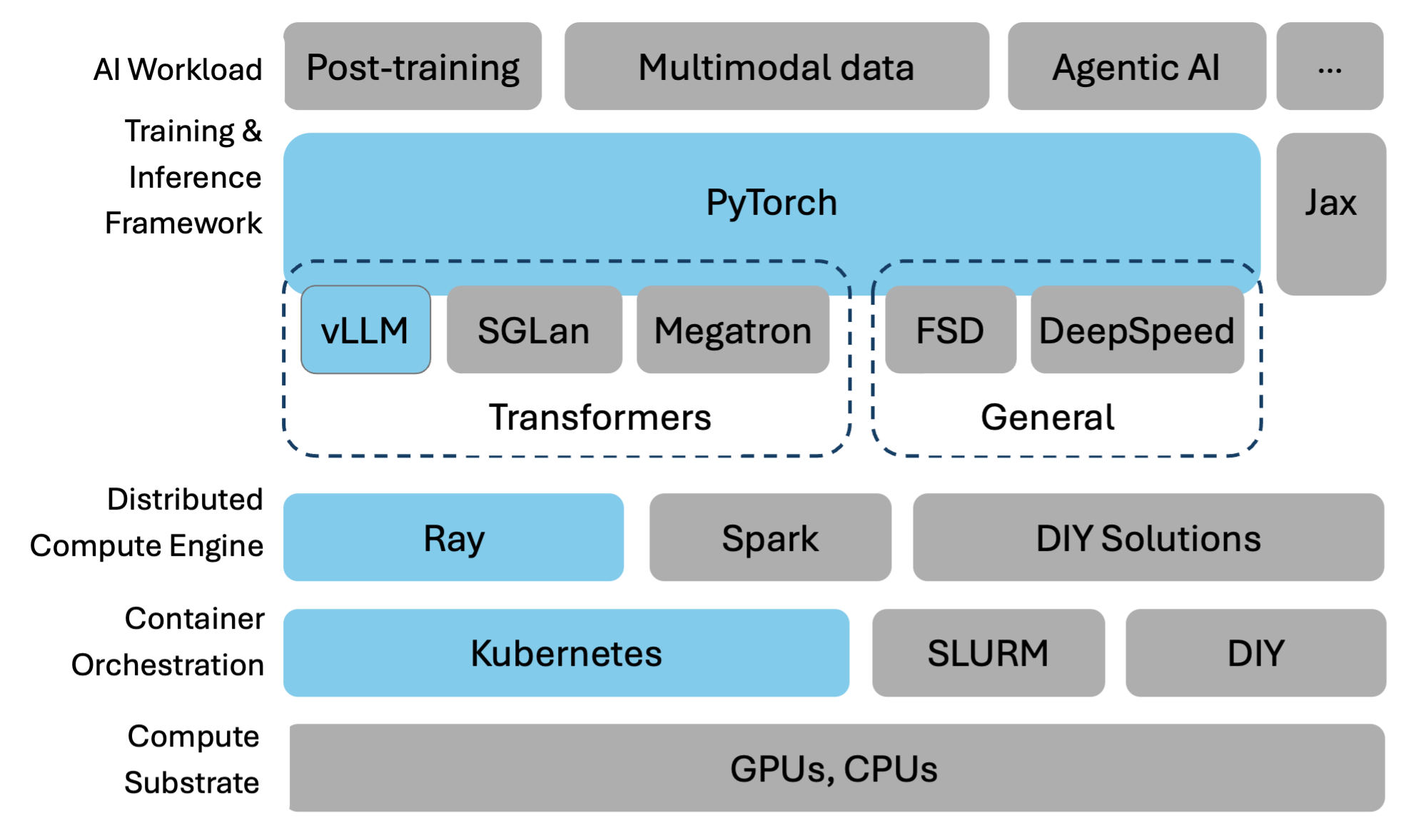}
    \caption{A layered architecture stack for AI infrastructure, comprising (bottom to top): a Compute Substrate of GPUs and CPUs; a Container Orchestrator, primarily Kubernetes; a Distributed Compute Engine; a Training \& Inference Framework spanning transformer-focused and general distributed training libraries; and an AI Workload layer. Figure reproduced from Anyscale~\cite{rayarch}.}
    \label{fig:rayarch}
\end{figure}

Building a modern inference platform for \glspl{LLM} requires a layered stack that balances high-performance model execution with robust infrastructure orchestration
. The gold standard architecture for this purpose comprises vLLM~\cite{kwon2023} as the \gls{LLM} engine, Ray~\cite{moritz2018ray} as the distributed runtime, and KubeRay~\cite{kanso2021} as the Kubernetes-native orchestrator. This architecture, illustrated in Figure \ref{fig:rayarch}, is garnering increasing interest from \gls{I-AI} users.


vLLM provides a state-of-the-art, low latency, high throughput, memory efficient engine for serving production-level \glspl{LLM}. It integrates natively with Ray and KubeRay and its features include: token generation; PagedAttention for memory optimized; distributed and coordinated \gls{KV} cache management; optimized hardware support for distributed inference techniques like tensor, pipeline and expert parallelism; and integration with popular open-source model repositories like Hugging Face~\cite{wolf2019huggingface}. 

Ray is an open-source distributed runtime framework that simplifies deployment and scaling of AI applications by providing features like complex model and workload distribution strategies, request routing, and autoscaling. It provides functionality for AI use cases expected by \gls{I-AI} users such as training and inference, and an abstract view of compute resources so that applications can be deployed declaratively, reducing the distributed-compute overhead of users. Ray integrates with Kubernetes through the KubeRay operator, which provides custom resources for cloud-native deployments and application management; including, \gls{RBAC}, Kubernetes network security features, and dynamic resource allocation including node and pod lifecycles.

The architecture addresses both production-level inference performance and enterprise-grade security controls for multi-tenant sensitive workloads. We demonstrate it is possible to fulfill user ambitions by deploying a multi-node, distributed, AI model service using KubeRay, Ray, and vLLM.


%% file: impl-versions-baseline.tex
The software versions used in this work, for the Kubernetes cluster baseline and the two use cases, are listed in Table~\ref{tab:versions}.

\begin{table}[ht]
  \small
  \centering
  \caption{Software Versions Baseline}
  \label{tab:versions}
  \begin{threeparttable}
      \begin{tabular}{lll}
        \toprule
        \textbf{Software} & \textbf{Version} \\
        \midrule
        HPE CSM Recipe & 25.3.3 \\
        \emph{libfabric} & 2.1.x\textdagger \\
        OpenMPI & 5.0.6 \\
        OSU Micro-Benchmarks & 7.5.2 \\
        CNI Plugins & v1.6.2 \\
        VNI Service & Commit 95869 \\
        RKE2 & v1.34.4+rke2r1 \\
        Metacontroller & v4.12.5 \\
        SmarterDeviceManager & 0.0.10 \\
        \midrule
        FRIDGE & Commit dbf74 \\
        Argo Workflows & 4.0.4 \\
        \midrule
        KubeRay & 1.5.1 \\
        Ray & 2.54.0 \\
        vLLM & 0.15.0 \\
        NCCL & 2.26.6 \\
        aws-ofi-nccl & 1.17.1 \\
        \bottomrule
      \end{tabular}
      \begin{tablenotes}
        \item[\textdagger Patched to support Slingshot-Kubernetes integration.] 
      \end{tablenotes}
  \end{threeparttable}
\end{table}

%% file: impl-VNI-isolation.tex
\subsection{Bare-metal Kubernetes and Multi-tenancy}
\label{section:baremetal_k8s_and_multi_tenancy}

On \gls{I-AI}, Kubernetes multi-tenancy requires Slingshot fabric network isolation to prevent workloads from interfering with one another. We implement the solution of Friese \emph{et al.}~\cite{friese2025closinghpccloudconvergencegap} including extension of the \gls{CXI} driver, \gls{CXI} library, and \emph{libfabric} to support creation of \gls{CXI} services scoped to Linux network namespaces; use of the \gls{CXI} \gls{CNI} plugin to provide Kubernetes' container-level network integration; and use of the \gls{VNI} service to allow Kubernetes to manage \gls{RDMA} endpoints.

In the solution~\cite{friese2025closinghpccloudconvergencegap}, users with workloads requiring Slingshot fabric isolation add a \gls{VNI} annotation to the metadata of their manifests. The \gls{VNI} service watches for annotated resources and creates \gls{VNI} \gls{CRD} reservations. Subsequently, when containers are created, the \gls{CXI} \gls{CNI} plugin is invoked, which checks the owning resource for a \gls{VNI} annotation and extracts the network namespace \emph{inode} of the container being constructed. The plugin then fetches the reserved \gls{VNI} and creates a \gls{CXI} service scoped to the network namespace and \gls{VNI}. In container deletion, the plugin tears down the \gls{CXI} service and \gls{VNI} service releases \gls{VNI} reservations that are no longer in use by a \gls{CXI} service.

Two \gls{VNI} allocation models are presented by Friese \emph{et al}. In the \emph{Per-Resource} model, each workload annotated with \yamlinline{vni: "true"} receives an exclusive \gls{VNI} \gls{CRD} instance providing isolation between different jobs. In the \emph{\gls{VNI} Claim} model, a named \emph{VniClaim} is first created independently. Each workload annotated with \yamlinline{vni: "<claim-name>"} is then granted access to the same \gls{VNI}, enabling communication between jobs. We note that the latter allocation model had not been implemented in the version of \gls{CXI} \gls{CNI} plugin available to us, and so we rely on the former.



In order to provision a Kubernetes cluster for this work, we isolate a group of compute nodes into a Slurm reservation that prevents user job scheduling. Subsequently, we deploy \gls{RKE2} with Longhorn~\cite{longhorn}, the NVIDIA GPU Operator~\cite{nvidiagpuoperator}, and the KubeRay Operator~\cite{kuberay} onto this node group. The \gls{CXI} \gls{CNI} chained plugin was compiled and added to Flannel following the procedure outlined by Friese \emph{et al.}~\cite{friese2025closinghpccloudconvergencegap}. We followed the procedure to deploy the \gls{VNI} service including: installing Metacontroller~\cite{metacontroller}, installing SmarterDeviceManager~\cite{smarterdevicemanager}, installing the \gls{CRD} and controller, compiling the endpoint binary with Golang, and building the endpoint image with Podman. All binaries and container images were built for ARM64 on \gls{I-AI}'s compute nodes themselves.

We modify the solution for \gls{I-AI}'s software environment and, notably, for its ARM64 architecture. Inside Kubernetes containers, the \emph{libfabric} \gls{CXI} provider discovery routine attempts to locate the \gls{HSN} \emph{netdev} associated with each \gls{CXI} character device to glean metadata required for initialisation e.g. \emph{hsn0} for \emph{cxi0}. However, the SmarterDeviceManager~\cite{smarterdevicemanager} used to mount the network devices does not mount the \gls{HSN} \emph{netdev} devices and the \emph{sysfs} entries are not visible inside the containers. This causes the \emph{netdev} lookup method to fail leaving \emph{libfabric} without usable interfaces and unable to initialise \gls{CXI} providers. We add a small patch to the \emph{netdev\_lookup()} method of \emph{libfabric} that reads two new environment variables, \emph{CXIP\_DEFAULT\_LINK} and \emph{CXIP\_DEFAULT\_SPEED}, to supply the link state and interface speed, respectively, that would normally be read from the \gls{HSN} \emph{netdev} \emph{sysfs} entries. In combination with \emph{CXIP\_SKIP\_AMA\_CHECK}, these changes enable the \gls{CXI} provider to successfully enumerate and open the \gls{CXI} character devices without requiring \emph{netdev} entries present in the container. Note, this patch is best understood as a pragmatic workaround rather than a desirable long-term solution. We also extend SmarterDeviceManager to allow \emph{gdrdrv} character device mounts to support GDRCopy inside containers. We discuss alternatives to these solutions and potential replacements for SmarterDeviceManager such as the \gls{HPE} \gls{CXI} Kubernetes Device Plugin~\cite{hpek8sdeviceplugin} and Akri~\cite{akri} in Section \ref{future}. 

The immutability of SquashFS precludes \emph{ad hoc} modifications to the booted \gls{SHS} stack. As such, the \gls{CXI} driver, \gls{CXI} library, and \emph{libfabric} modifications are baked into a new image artifact and deployed with \gls{CSM} onto our node group. Principally, the \gls{SHS} \gls{CFS} layer was changed to point toward a Nexus repository containing the patched \gls{SHS} RPMs.

Figure \ref{fig:vnicode} represents a minimal Kubernetes manifest containing the necessary modifications for the \emph{Per-Resource} \gls{VNI} allocation model, including the \gls{VNI} annotation, \gls{CXI} device mount, host library mounts, and environment variables.

\begin{figure}[ht]
\Description{A Kubernetes manifest illustrating the changes required to implement Linux namespace-based VNI isolation. Notably, the addition of "vni: true", inclusion of CXIP_DEFAULT_LINK, CXIP_DEFAULT_SPEED, CXIP_SKIP_AMA_CHECK environment variables, and inclusion of \emph{libcxi} and \emph{libfabric} mounts from the host.}
\begin{lstlisting}[style=myyaml]
apiVersion: apps/v1
kind: Deployment
metadata:
  annotations:
    vni: "true"
  labels:
    app: demo
  name: demo
spec:
  replicas: 2
  selector:
    matchLabels:
      app: demo
  strategy: {}
  template:
    metadata:
      labels:
        app: demo
    spec:
      containers:
      - image: busybox
        name: busybox
        command: ["sleep","infinity"]
        env:
        - name: LD_LIBRARY_PATH
          value: "/usr/lib64/:/extras/lib/:/hostlibs"
        - name: CXIP_DEFAULT_LINK
          value: "1"
        - name: CXIP_DEFAULT_SPEED
          value: "200000"
        - name: CXIP_SKIP_AMA_CHECK
          value: "1"
        resources:
          requests:
            smarter-devices/cxi0: "1"
          limits:
            smarter-devices/cxi0: "1"
        volumeMounts:
        - name: hostlibs
        mountPath: /hostlibs
        - name: libfabric
        mountPath: /extras
    volumes:
    - name: hostlibs
      hostPath:
        path: /usr/lib64/
        type: Directory
    - name: libfabric
      hostPath:
        path: /opt/shs-libfabric/build
        type: Directory
\end{lstlisting}
\caption{Minimum example of a Kubernetes deployment manifest for testing network namespace-based Slingshot fabric isolation between containers. The \gls{CXI} device, environment variables, and libraries required to execute \emph{fi\_pingpong} tests are provided to the containers.}
\label{fig:vnicode}
\end{figure}

%% file: impl-FRIDGE.tex
\subsection{Trusted Research Environments}

We deploy a \gls{FRIDGE} instance on \gls{I-AI} using the Python implementation of Pulumi infrastructure-as-code~\cite{pulumi}, which is publicly available~\cite{watson2025code}. The Pulumi code provisions all Kubernetes namespaces, \gls{RBAC} bindings, applications, Custom Resource Definitions, and network policies required for a \gls{FRIDGE} instance.

\gls{FRIDGE} requires load balancer IPs for its API and internal services. On \gls{I-AI}, \gls{BGP} peering is not readily available on either Rosetta or edge switches, precluding standard \gls{BGP}-based load balancer addresses using Layer~2 \gls{ARP} announcement. Modifications to these switches are not desirable on a production system. We therefore configure MetalLB~\cite{metallb} to advertise load balancer addresses using Layer~2 \gls{ARP} announcement. Container networking is provided by Cilium~\cite{cilium}, which enforces per-namespace \emph{NetworkPolicies} to ensure pod traffic within on \gls{TRE} tenancy cannot reach pods from another. In future work, we will investigate dedicated load balancer nodes hosted outside of the Slingshot fabric to enable \gls{BGP} support. 

Since \gls{RDMA} traffic does not traverse the standard Linux kernel networking stack, conventional network policies have limited effect on traffic flows. By ensuring that workloads intending to use \gls{RDMA}, specifically those that mount \gls{CXI} devices, are assigned \gls{CXI} services with dedicated \glspl{VNI} we can guarantee that such workloads remain isolated from the rest of the Slingshot fabric.

Longhorn~\cite{longhorn} is used in combination with node-local SSDs to provide encrypted volumes to meet the requirement of data encryption at rest.

\gls{FRIDGE} uses Argo Workflows~\cite{argoworkflows} for job orchestration. We provide \emph{WorkflowTemplates} pre-configured to mount the \gls{CXI} devices and host library dependencies described in Section \ref{section:baremetal_k8s_and_multi_tenancy}, ensuring that all jobs submitted through the \gls{FRIDGE} API have access to the Slingshot fabric without requiring manual low-level interventions from users. The \emph{WorkflowTemplate} metadata carries the \yamlinline{vni: "true"} annotation, which is inherited by each \emph{Workflow} it instantiates. We also update the Metacontroller \emph{DecaratorController}~\cite{metacontroller} of the \gls{VNI} service to look for \emph{Workflow} creation events, as shown in Figure \ref{fig:decaratorcontrollercode}. Each job submitted through the \gls{FRIDGE} API is then guaranteed to have all its pods given the same \gls{VNI} allocation. 

\begin{figure}[ht]
\Description{Modifications to the Metacontroller \emph{DecaratorController} of the \gls{VNI} service that enable it to listen for Argo \emph{Workflow} and \emph{RayCluster} creation events for the purposes of \gls{VNI} allocation.}
\begin{lstlisting}[style=myyaml]
apiVersion: v1
kind: List
items:
- apiVersion: metacontroller.k8s.io/v1alpha1
  kind: DecoratorController
  metadata:
    name: vni-autovni-controller
  spec:
    attachments:
    - apiVersion: horizon-opencube.eu/v1
      resource: vnis
    hooks:
      finalize:
        version: v1
        webhook:
          url: http://vni-endpoint-service.vni-management:8842/finalize
      sync:
        version: v1
        webhook:
          url: http://vni-endpoint-service.vni-management:8842/sync
    resources:
    - annotationSelector:
        matchExpressions:
        - key: vni
          operator: Exists
      apiVersion: apps/v1
      resource: deployments
    #---snip---
    - annotationSelector:
        matchExpressions:
        - key: vni
          operator: Exists
      apiVersion: argoproj.io/v1alpha1
      resource: workflows
    - annotationSelector:
        matchExpressions:
        - key: vni
          operator: Exists
      apiVersion: ray.io/v1
      resource: rayclusters
    - apiVersion: horizon-opencube.eu/v1
      resource: vniclaims
\end{lstlisting}
\caption{Modifications to the Metacontroller \emph{DecaratorController} of the \gls{VNI} service that enable it to listen for Argo \emph{Workflow} and \emph{RayCluster} creation events for the purposes of \gls{VNI} allocation.}
\label{fig:decaratorcontrollercode}
\end{figure}

The \emph{argo-workflows} namespace is configured with the Kubernetes privileged pod security standard to permit the host library mounts required for \gls{CXI} access. This is a broad grant and we explore less intrusive solutions in Section \ref{sec:summary}.

%% file: impl-model-hosting.tex
\subsection{Secure and Multi-node Model Hosting}
\label{sec:implementationmodel}
We deploy the KubeRay, Ray, and vLLM stack to show the multi-node AI sandboxing use case discussed in section \ref{sec:background}.

Ray provides an ARM64 compatible container image with functionality for cluster configuration and management which is readily deployable within Kubernetes. The Ray base image does not contain components for vLLM inference or ARM64 and CUDA compatible PyTorch libraries. The Ray documentation recommends adding dependencies at runtime by applying a \emph{runtime\_env} flag to the application configuration which encapsulates packages and installation instructions. This is unsuitable in our deployment for two reasons: firstly, because the required libraries are large and would add unreasonable overhead to pod initialisation; secondly, the increased startup overhead would require changing Kubernetes timeout configurations so that pods are not prematurely put into unusable states by Kubernetes cluster management.

To enable deployment on \gls{I-AI} we construct a multi-stage custom image build pipeline that incorporates fabric-specific and inference dependencies at image build time. This pipeline has added benefit of reduing reliance on a compatible host operating system and isolates software within the container. The first stage produces a fabric-aware base image. Starting from the official Ray \texttt{2.54.0-py312-cu126-aarch64} image~\cite{raydocker}, a Spack~\cite{spack,10.1145/2807591.2807623} environment is constructed containing the following components, compiled for CUDA compute architecture \texttt{sm\_90}: \emph{libfabric} with \emph{netns}-aware \gls{CXI} provider enabled, built with CUDA-aware and GDRCopy-enabled options, and linked against \emph{netns}-aware \emph{libcxi} and \emph{cxi-driver}; \gls{NCCL} with CUDA support; aws-ofi-nccl transport plugin; OpenMPI compiled with \gls{OFI} support. The second stage builds the inference image on top of the first stage image. PyTorch with CUDA support is installed from the PyTorch binary wheel index. vLLM is then installed alongside \emph{mistral-common} for Mistral tokenizer support. Within the resultant image, Ray can be executed after the Spack environment is sourced and activated.

KubeRay provides two patterns for application deployment: Imperative using \emph{RayJob} and Declarative using \emph{RayService}. We adopt the declarative approach for cluster and application deployment as it enables user control over cluster creation and scaling; continuous, persistent and reliable inference endpoints guaranteed by Kubernetes lifecycle management; and the ability to declare application configurations using patterns and syntax familiar to users.

The declarative approach is recommended by Ray for use cases that require persistence and reliability guarantees. The model sandboxing use case aligns with this because it requires persistent hosting. The declarative approach also conforms to the \gls{CXI} provision discussed in section \ref{section:baremetal_k8s_and_multi_tenancy}, as it allows libraries and environment variables to be declared within the application configuration in the same way shown in Figure \ref{fig:vnicode}. This can be achieved at vLLM application runtime using the \emph{runtime\_env} deployment flag mentioned above, or at pod initialisation time using Kubernetes-native \emph{volume} and \emph{volumeMount} components. Using this approach, we deploy a Kubernetes manifest with Ray head and worker pods as required, and with the pre-cached model weights injected via a \emph{volumeMount}.

\begin{figure}[ht]
\Description{Mutating Admission Policy code to annotate \emph{RayCluster} resources with \yamlinline{vni: "true"} in order for \gls{VNI} allocation.}
\begin{lstlisting}[style=myyaml]
apiVersion: admissionregistration.k8s.io/v1beta1
kind: MutatingAdmissionPolicy
metadata:
  name: add-raycluster-annotations
spec:
  reinvocationPolicy: IfNeeded
  failurePolicy: Fail
  matchConstraints:
    resourceRules:
      - apiGroups: ["ray.io"]
        apiVersions: ["v1"]
        operations: ["CREATE"]
        resources: ["rayclusters"]
  matchConditions:
    - name: has-no-annotation
      expression: "!has(object.metadata.annotations) || !('vni' in object.metadata.annotations)"
  mutations:
    - patchType: JSONPatch
      jsonPatch:
        expression: >
          [
            JSONPatch{
              op: "add",
              path: "/metadata/annotations/vni",
              value: "true"
            }
          ]
---
apiVersion: admissionregistration.k8s.io/v1beta1
kind: MutatingAdmissionPolicyBinding
metadata:
  name: add-raycluster-annotations-binding
spec:
  policyName: add-raycluster-annotations
\end{lstlisting}
\caption{Mutating Admission Policy code to annotate \emph{RayCluster} resources with \yamlinline{vni: "true"} in order for the \emph{RayCluster} to receive \gls{VNI} allocations.}
\label{fig:mapcode}
\end{figure}

\gls{VNI} allocation requires top-level resource owners to be annotated with \yamlinline{vni: "true"}. In our implementation this is the \emph{RayCluster}. The KubeRay declarative pattern creates the \emph{RayCluster} dynamically on initialization of each \emph{RayService}, we use a combination of \gls{MAP}~\cite{map} (see Figure \ref{fig:mapcode}) and the Metacontroller \emph{DecaratorController} to watch for \emph{RayCluster} creation events (see Figure \ref{fig:decaratorcontrollercode}). Each pod spawned under the \emph{RayCluster} triggers the CXI CNI chained plugin workflow which checks the pod \emph{ownerReferences} to find the parent VNI custom resource and create a \emph{CXI service} for the pod’s \emph{netns} with the parent VNI. The mechanism of claiming a CXI service ID for each pod under the \emph{RayCluster} leads to a resource bottleneck in the underlying Cassini hardware because each CXI service claims a Triggered List Entry pool of which there are only three available to users per each \gls{NIC}.

%% file: eval-K8s.tex
\subsection{Bare-metal Kubernetes and Multi-tenancy}

We have successfully reproduced the Slingshot namespace-based network isolation implementation of Friese \emph{et al.} on \gls{I-AI}'s ARM64 architecture and validated it through a series of confidence tests using \emph{fi\_pingpong}. First, we confirmed baseline fabric connectivity between two containers on the same host, each with a distinct \gls{CXI} service scoped to their respective network namespace operating with the same \gls{VNI}. Second, we confirmed the same connectivity between two containers on different host nodes. Third, we verified \gls{VNI} isolation enforcement by manually adjusting the \gls{VNI} on one side of the interaction and observing expected communication loss. Finally, as a preliminary penetration test, we attempted to use a third container to snoop on traffic between two containers by supplying it with one of the victim's \gls{CXI} service IDs and \glspl{VNI}. As expected, endpoint creation fails because the malicious container does not possess the correct namespace. 

\begin{figure}[htbp]
\Description{Two line charts showing OSU Micro-Benchmark results for inter-node point-to-point communication over the Slingshot \gls{CXI} fabric. Chart (a) shows bandwidth in MB/s against message size in bytes on a logarithmic scale. Chart (b) shows latency in microseconds against message size in bytes on a logarithmic scale. Both charts contain three lines: bare-metal, Kubernetes without \gls{VNI} isolation, and Kubernetes with \gls{VNI} isolation. In both charts the two Kubernetes lines are indistinguishable from one another across all message sizes. However, both Kubernetes lines differ from the bare-metal baseline by a fixed offset of increased latency and decreased bandwidth, which is only visible for small messages.}
    \centering
    \begin{subfigure}[b]{0.48\textwidth}
        \centering
        \includegraphics[width=\textwidth]{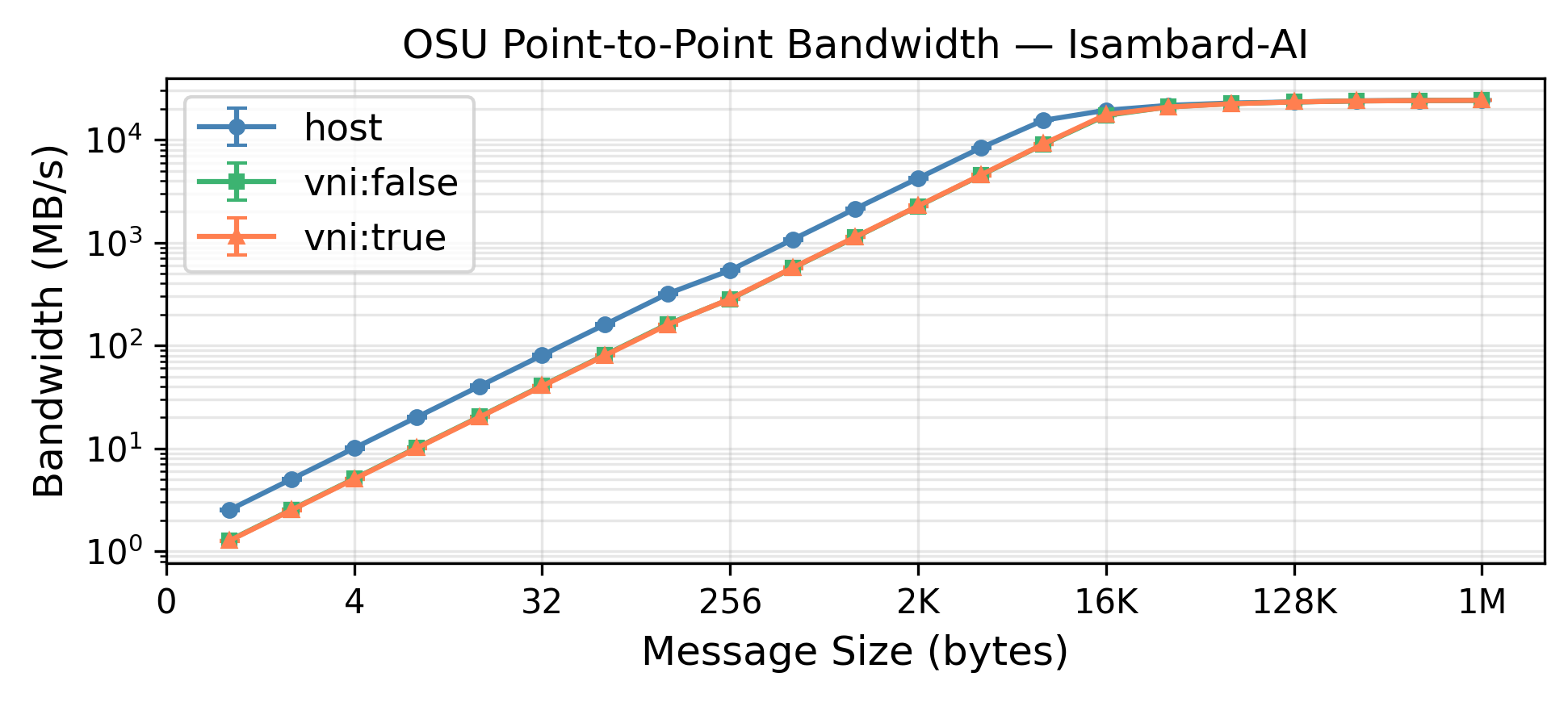}
        \caption{Point-to-point bandwidth (\texttt{\emph{osu\_bw}})}
        \label{fig:osubw}
    \end{subfigure}
    \hfill
    \begin{subfigure}[b]{0.48\textwidth}
        \centering
        \includegraphics[width=\textwidth]{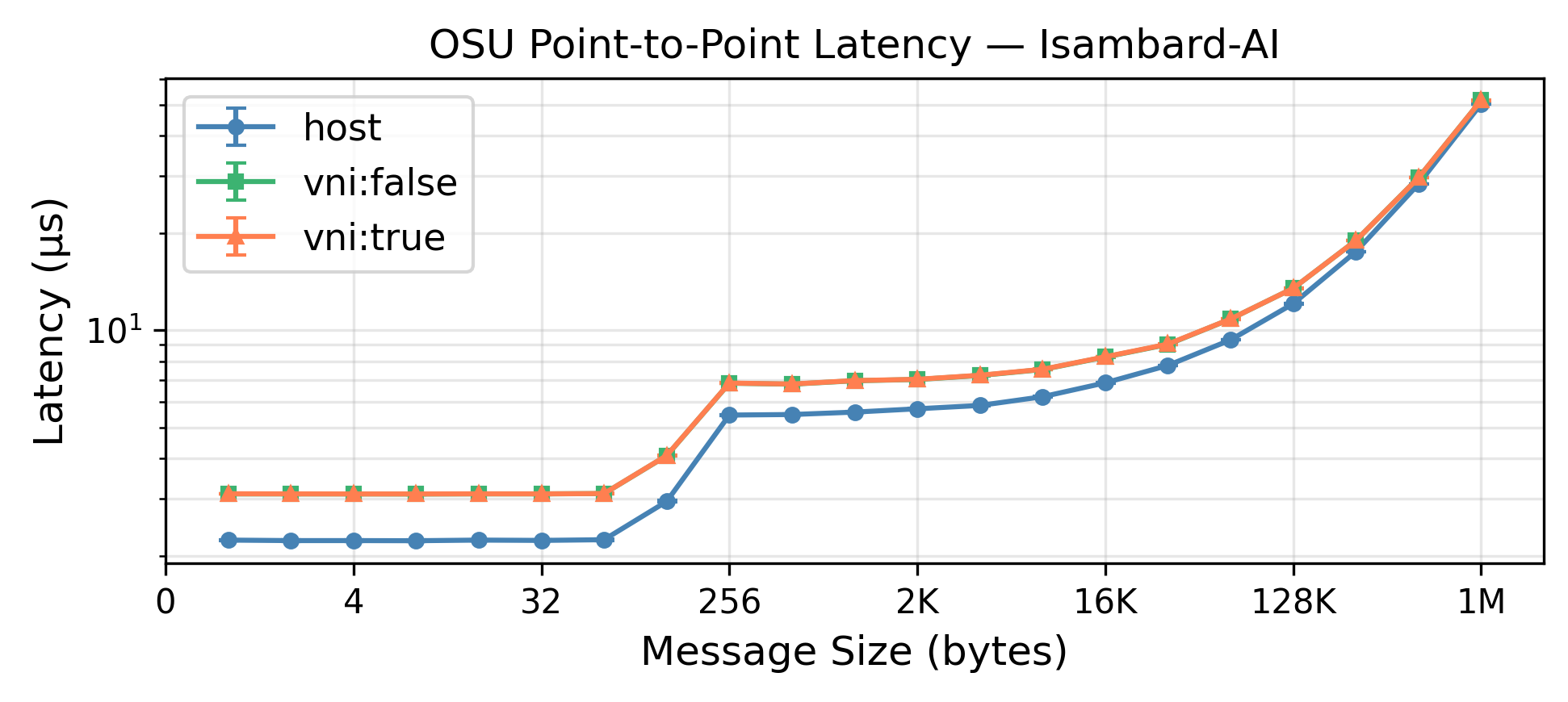}
        \caption{Point-to-point latency (\texttt{\emph{osu\_latency}})}
        \label{fig:osulat}
    \end{subfigure}
    \caption{OSU Micro-Benchmarks inter-node point-to-point (a) bandwidth and (b) latency on bare-metal, Kubernetes without network namespace-based \gls{VNI} isolation, and Kubernetes with Linux network namespace-based VNI isolation.}
\end{figure}

Figures \ref{fig:osubw} and \ref{fig:osulat} show OSU Micro-Benchmarks~\cite{osu} point-to-point bandwidth (\emph{osu\_bw}) and latency (\emph{osu\_latency}) results for three different test configurations; bare-metal, and Kubernetes containers with and without network namespace-based \gls{VNI} isolation, between different nodes. We average 10 runs per test using 10,000 and 20,000 for the iteration counts for bandwidth and latency measurements, respectively. Both Kubernetes configurations, with and without namespace-based \gls{VNI} isolation, express statistically identical latency profiles that converge to \textasciitilde3.11$\mu$s for small messages. We conclude that namespace-based \gls{VNI} isolation adds no meaningful latency penalty. These latency profiles carry a fixed overhead of \textasciitilde0.89$\mu$s compared to those observed for both bare-metal and Podman containers launched directly via Slurm, which both converge to \textasciitilde2.22$\mu$s for small messages. We attribute this latency overhead to Kubernetes network namespace boundary traversal on the data path. The latency overhead manifests a similar reduction in bandwidth for small messages in Kubernetes environments, but in all cases the peak bandwidth converges to the theoretical maximum of the Slingshot fabric of 200Gb/s as message sizes increase. These fixed overheads do not scale with message size and so are unlikely to significantly affect throughput-bound workloads. Moreover, factors like rendezvous protocol parameters play far more of a role in performance, which should be noted by users of the system.


%% file: eval-FRIDGE.tex
\subsection{Trusted Research Environments}


We evaluate the \gls{FRIDGE} deployment, executed over the Slingshot-isolated Kubernetes stack described in Section \ref{sec:implementation}, using a representative Argo Workflow submitted by a \gls{TRE} researcher. The \emph{WorkflowTemplate} encapsulates the necessary resources; mounts, libraries, environment variables, and annotations; required for a workload to securely use the Slingshot network. This allows users to submit jobs without managing low-level \gls{CXI} configuration directly. 

A key operational concern is workflow lifetime; Argo \emph{Workflows} are not automatically deleted after their pods have completed. In our deployment, \gls{VNI} reservations are only released once the owning \emph{Workflow} custom resource is deleted. This gradually leads to exhaustion of \glspl{VNI} from the available pool. Therefore, we enforce a \emph{ttlStrategy} in the template configuration. Moving forward, we will investigate the Workflow Archive to maintain auditability.

We also identify several issues specific to the bare-metal Slingshot networking stack. Cilium and MetalLB provide the container networking and load-balancer functionality required by \gls{FRIDGE}. Cilium's default handling of \gls{ARP} replies interfered with routing on \gls{HSN} interfaces, which carry CIDR-based Slingshot configurations rather than conventional host IPs, preventing access to the Kube \gls{API} from outside the nodes themselves. We resolve this by manually enabling \gls{ARP} replies on the interfaces and assigning an address on the \gls{HSN} to the default-route interface. These issues stem from hosting the control-plane and Slingshot workload traffic on the same nodes and could be addressed by decoupling the control-plane from the Slingshot fabric and providing external load-balancer IPs. The \gls{CXI} \gls{CNI} plugin itself required only minimal changes to operate alongside Flannel and Cilium, suggesting the approach is portable across different \gls{CNI} providers.

Storage integration was comparatively simple. Longhorn required minor adjustments to work on \gls{I-AI} with local node storage, and was sufficient for demonstration and evaluation workloads. A limitation of this approach is that storage capacity scales only with number of nodes in the isolated \gls{FRIDGE} cluster. Future work will investigate scalable shared multi-tenant storage.

\gls{FRIDGE} relies on Kubernetes pod security standards, but workloads requiring Slingshot access currently execute in a privileged namespace. This presents a broad attack surface and undermines the intent of the isolation model of the cluster. A more secure approach could involve workload controls through runtime classes, Pod Security Admissions, Validating Admission Policies, or with tools like Kyverno~\cite{kyverno} and KubeArmour~\cite{kubearmor}.

%% file: eval-model-hosting.tex
\subsection{Secure and Multi-node Model Hosting}

We evaluate multi-node LLM serving performance for models deployed to the \gls{FRIDGE} cluster with Slingshot namespace \gls{VNI} isolation as the communication layer between nodes. The AI sandboxing use case deployed to Kubernetes allows for pipeline agnostic deployment of models, meaning model pipelines can be split between cluster compute components in a way that flexibly allocates resources. We measure the inter-node communication overhead from the integration of \gls{VNI} isolation with Slingshot. The results presented here are not for the purpose of evaluating LLM serving optimization, but to show that our AI sandboxing stack does not degrade user experience and performance.

First, as a confidence test, we measure the \gls{TTFT} and \gls{TPS}~\cite{Bento_ML:LLM_metrics} with a custom Python benchmarking script using the \emph{requests} library to directly call the API endpoint on the deployment without the Slingshot namespace-based \gls{VNI} isolation. The script uses a single streaming request to the endpoint and records the time elapsed between request and first response using the \emph{time} module. Subsequent tokens are captured and counted to obtain a \gls{TPS} estimate. Prior to measurement requests, a warmup request is issued to prime the model. We measure the average \gls{TTFT} and \gls{TPS} to be 120ms and 22, respectively, over 10 runs each with a batch size of one with the model distributed between two nodes. The \gls{TTFT} compares favourably to that of 240ms reported by Cerebras for the same model on their wafer hardware~\cite{cerebras2024llama405b}. The \gls{TPS} is comparable to 36 \gls{TPS} reported by Oracle for the same model running on eight AMD MI300X GPUs with a batch size of one~\cite{oracle2024mi300x}.

For a comprehensive evaluation of the AI sandboxing stack we use the vLLM bench framework~\cite{vLLM:benchmark_CLI}. It allows simulation of real-world scenarios for deployed models. The user specifies the model's API endpoint along with a set of parameters that simulates the expected workload and a dataset of sample prompts. In order to simplify benchmarking, we maintain vLLM default settings for a number of parameters. For example, the layer split strategy is set to evenly split model layers between GPUs and nodes. GPU utilization is set to the default 0.9. We change the \emph{-{}-request-rate} parameter to 20 requests per second, this is recommended by vLLM when attempting to simulate realistic traffic patterns. We also set the \emph{-{}-max-ongoing-requests} parameter in Ray to 128 to limit queuing overhead. 

The metrics of greatest interest are; Output \gls{TPS} to measure system capacity, and \gls{TPOT} and \gls{ITL} to measure responsiveness. Output \gls{TPS} is a measure of the models token generation throughput across all active requests, \gls{TPOT} is the average time between generating output tokens excluding the first, while \gls{ITL} refers to the average time between each consecutive token generated within a sequence.

\begin{figure*}[!htbp]
    \centering
    \begin{subfigure}[b]{0.33\textwidth}
         \centering
         \includegraphics[width=\textwidth, height=0.75\textwidth]{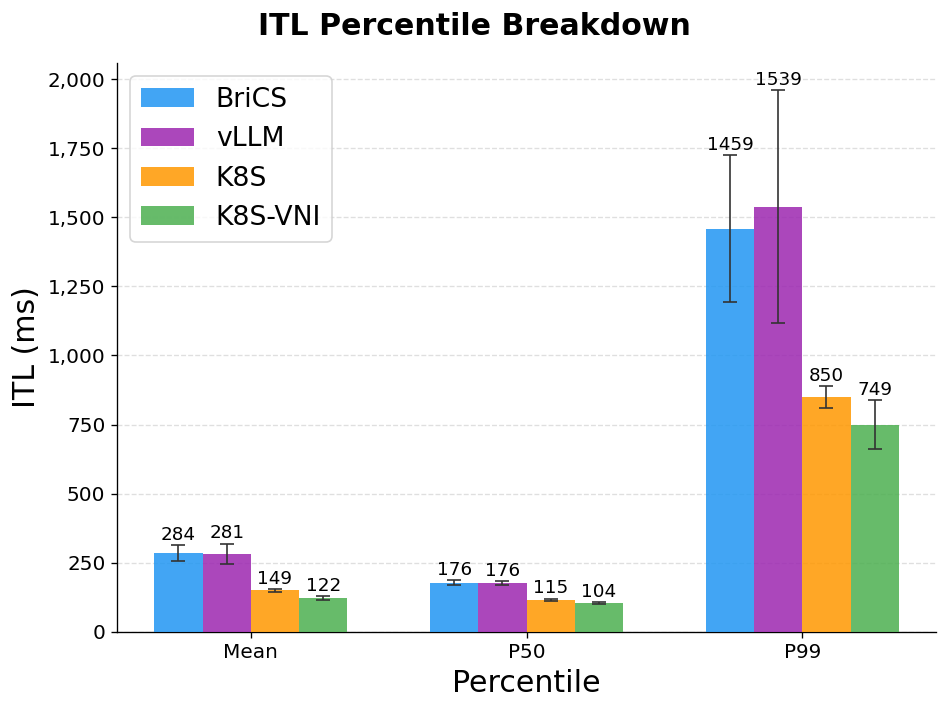}
         \caption{Mean ITL}
     \end{subfigure}
     \begin{subfigure}[b]{0.33\textwidth}
         \centering
         \includegraphics[width=\textwidth, height=0.75\textwidth]{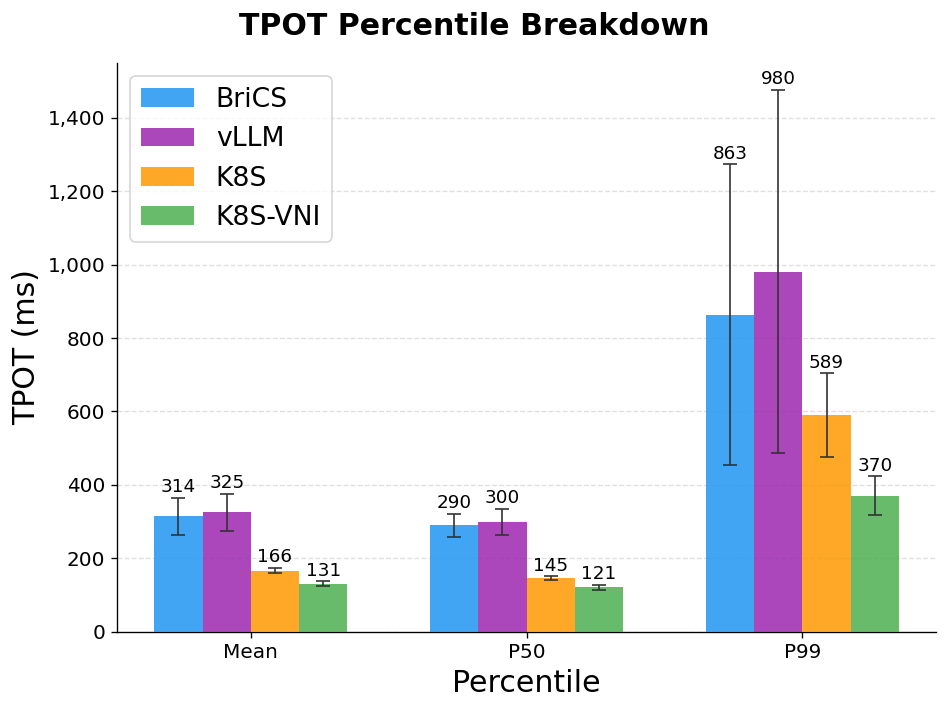}
         \caption{Mean TPOT}
     \end{subfigure}
     \begin{subfigure}[b]{0.33\textwidth}
         \centering
         \includegraphics[width=\textwidth, height=0.75\textwidth]{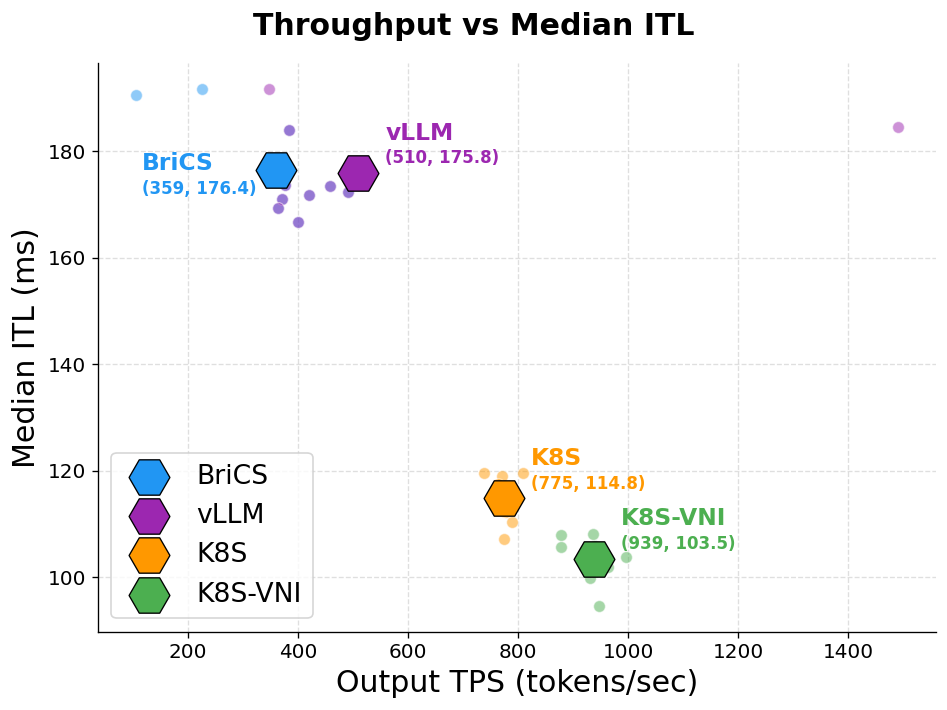}
         \caption{Output TPS by Median ITL}
     \end{subfigure}
    \caption{Benchmarking results on the \emph{ShareGPT\_Vicuna\_unfiltered} dataset for the \emph{Meta-Llama-3.1-405B-Instruct-FP8} model deployed in the four infrastructure setups. (a) and (b) show percentile performance for the \gls{ITL} and \gls{TPOT} metrics, a lower time in milliseconds (\emph{ms}) is better. P99 refers to he value below which 99\% of requests fall, it reveals worst-case performance for the slowest 1\% of requests. (c) shows the relationship between Output \gls{TPS} and median \gls{ITL}, a higher \gls{TPS} is better.}
    \label{fig:vllm_serve_results}
\end{figure*}

We use the \emph{ShareGPT\_Vicuna\_unfiltered} dataset which contains 53K example conversations from user prompts to ChatGPT~\cite{anon8231489123_sharegpt}. We deploy the \emph{Meta-Llama-3.1-405B-Instruct-FP8}\cite{metallama} model in a \emph{tensor\_parallel=4}, \emph{pipeline\_parallel=2} configuration. Meaning the model is deployed using two \gls{I-AI} nodes, utilizing 4 GPUs on each node.

We compare four separate infrastructure configurations; the \gls{BriCS} guidelines on how to perform distributed vLLM inference on \gls{I-AI}~\cite{brics_distributed_vllm_inference}, the vLLM tutorial on using Ray for multi-node inference~\cite{vllm_distributed_multi_node_inference}, along with our AI Sandboxing stack with the \gls{VNI} component turned \emph{off} and \emph{on}. We perform 10 runs of vLLM bench for each of our hardware configuration setups.

Figure \ref{fig:vllm_serve_results} shows the performance of each of the hardware configurations we evaluate across the \gls{ITL}, \gls{TPOT} and Output \gls{TPS} metrics. The two setups that use our stack perform favorably compared to the \gls{BriCS} and vLLM alternatives. For both \gls{ITL} and \gls{TPOT} the \yamlinline{vni: "true"} setup achieves mean \emph{(ms)} performance of less-than half the \gls{BriCS} and vLLM setups. The performance improvement of the \gls{k8s} and \gls{VNI} deployments are also shown in output \gls{TPS} where the \yamlinline{vni: "true"} setup achieves approximately a \gls{TPS} speed-up of three times over the \gls{BriCS} configuration.

We observe that although our work does not concentrate on LLM serving optimization, we have inadvertently optimized the performance of our \gls{k8s} and \gls{VNI} deployment configurations. We followed the provided setup steps for both the \gls{BriCS} and vLLM configurations, which includes optimizations for \gls{NCCL} and \gls{RDMA}, and for \gls{BriCS} optimizations for Slingshot and \gls{I-AI}.

%% file: summ-conc.tex

Using real world use cases that represent the rapidly evolving secure model hosting and sensitive data environments we have demonstrated a reproducible approach for deploying multi-tenant, bare-metal Kubernetes platforms on \gls{I-AI}. Another important aspect that we confirmed is portability of cloud-native architecture and playbooks on \gls{I-AI}'s unique hardware profile, which uses Cray \gls{HPE} EX with NVIDIA Grace-Hopper GH200 and Slingshot interconnect, for both use cases. In other words, supercomputing platforms built on Ethernet-compatible interconnects such as Slingshot and cloud-like provisioning stacks such as \gls{CSM} can support multi-tenant services without requiring fully bespoke, vertically integrated solutions. However, several challenges remain that we would like to continue working on for hardening the solutions and for reducing the complexity for scale-out operations where we would welcome contributions and feedback from the community.

%% file: summ-future.tex

The current Slingshot isolation stack is function, but operationally complex. It relies upon several independent and loosely-coupled components including; Metacontroller, SmarterDeviceManager, the \gls{VNI} service, and the chained \gls{CNI} \gls{CXI} plugin. Many of these components have been subject to custom modifications as documented in Section \ref{sec:implementation}. Also, workloads requiring Slingshot access require several manifest augmentations, including \gls{VNI} annotations, hosts library and driver mounts, and environment variables describing device metadata. Reducing this operational burden for both administrators and end users is therefore an important direction for future work. A more robust solution would involve simplified implementation details that are elided from the user as much as possible. We outline suggestions and areas of future work to achieve this.

The \gls{CXI} driver, \gls{CXI} library, and \emph{libfabric} patches supplied by Friese \emph{et al}. are being upstreamed into \gls{SHS} version 14 release candidates, which will hopefully reduce administration overhead of future Slingshot isolation solutions. The supplemental patch to \emph{libfabric} introduced in this paper should be superseded by mounting \gls{CXI} character devices along with their associated \gls{HSN} \emph{netdev} devices. SmarterDeviceManager, originally developed for agricultural IoT applications, has not seen project activity in more than a year and is likely no longer maintained. This makes it difficult to request or implement feature and security updates. Future support for \emph{netdev} mounts is therefore unlikely. Akri~\cite{akri} emerges from the IoT space and presents a potential candidate replacement for device handling. It is actively maintained with a large community and supports both device discovery by IP and by \emph{udev} rules. A custom discovery handler could mount the \gls{CXI} character devices, \emph{gdrdrv} character devices, and \gls{HSN} \emph{netdev} devices as part of a single resource request. \gls{CXI} Kubernetes \gls{DRA} driver presents another viable candidate to address many of these problems. With some development, a purpose-built \gls{DRA} driver for \gls{CXI} devices could allow implicit \emph{netdev} discovery and mounting, as well as automatic mounting host dependencies. This approach could also eliminate the need for some of the modifications to components, like to those of Metacontroller to allow listening for custom resource creation.

The \gls{CNI} \gls{CXI} plugin and the \gls{VNI} service represent a robust solution and may benefit from minor optimisations. The \gls{CNI} \gls{CXI} plugin currently only supports direct parent resource discovery and does not traverse the ownership chain in its entirety. In Section \ref{sec:implementationmodel}, we present Kubernetes \glspl{MAP} as a solution to address this problem. \glspl{MAP} could be leveraged much more heavily to either warn about missing manifest parameters or to inject them automatically, providing a much more seamless user experience. As discussed in section \ref{sec:implementationmodel} the plugin approach triggers resource constraints caused by hardcoded limits in the underlying Cassini hardware. This limit could be overcome by removing the one-to-one dependency between VNI enabled pods and allocated \gls{CXI} services. This is feasible because individual \gls{CXI} services have the capacity for managing multiple \glspl{VNI}, however it would require more work to understand how AI applications utilize the \gls{NIC} resources allocated to each \gls{CXI}.

The \gls{VNI} service could also be modified to represent \glspl{VNI} as a cluster-wide \emph{ResourceSlice}. This would couple the \gls{VNI} lifecycle of workloads more natively into Kubernetes and avoiding the issue of \gls{VNI} exhaustion by granting the scheduler greater visibility into the \gls{CXI} allocation stack. Also, the \gls{VNI} service currently tracks \glspl{VNI} using an SQLite database, which could potentially be replaced with Custom Resources metadata to improve Kubernetes integration. Alternatively, the service could be integrated into the \gls{DRC2} \gls{VNI} allocation service provided by \gls{USS}~\cite{uss}.

The AI sandboxing use case could be extended to cover the full lifecycle of AI model development, especially relevant is the fine-tuning use case in which models are finetuned on data that requires the governance and access security restrictions provided by our deployment stack. The AI user experience could be improved by extending \gls{FRIDGE} and \gls{VNI} isolation to other AI frameworks such as Kubeflow for finetuning~\cite{kubeflow2026} and KServe for inference~\cite{kserve2026}. This would involve finding a new way to assign \gls{VNI} information to applications because they have different primitives and abstractions around how they treat model deployments.

%% file: main.bbl
\begin{thebibliography}{48}
\providecommand{\natexlab}[1]{#1}
\providecommand{\url}[1]{\texttt{#1}}
\expandafter\ifx\csname urlstyle\endcsname\relax
  \providecommand{\doi}[1]{doi: #1}\else
  \providecommand{\doi}{doi: \begingroup \urlstyle{rm}\Url}\fi

\bibitem[Sched{MD}(2026{\natexlab{a}})]{slinky}
Sched{MD}.
\newblock Slinky: Sched{MD}'s set of projects to enable interoperability between {S}lurm and {K}ubernetes, 2026{\natexlab{a}}.
\newblock URL \url{https://slurm.schedmd.com/slinky.html}.
\newblock [Online; accessed {M}arch 16, 2026].

\bibitem[Sched{MD}(2026{\natexlab{b}})]{kubeoperator}
Sched{MD}.
\newblock Kubernetes {O}perator for {S}lurm {C}lusters, 2026{\natexlab{b}}.
\newblock URL \url{https://slinky.schedmd.com/projects/slurm-operator/en/release-1.0/}.
\newblock [Online; accessed March 16, 2026].

\bibitem[CNCF(2026)]{cncfbatch}
CNCF.
\newblock Batch {S}ystem {I}nitiative {W}orking {G}roup, 2026.
\newblock URL \url{https://tag-runtime.cncf.io/wgs/bsi/}.
\newblock [Online; accessed March 16, 2026].

\bibitem[Volcano(2026)]{volcano}
Volcano.
\newblock Cloud-native {C}ontainer {B}atch {S}cheduler for {HPC} {W}orkloads, 2026.
\newblock URL \url{https://volcano.sh/en/docs/}.
\newblock [Online; accessed March 16, 2026].

\bibitem[Armada(2026)]{armada}
Armada.
\newblock Multi-{K}ubernetes {C}luster {B}atch {J}ob {M}eta-{S}cheduler, 2026.
\newblock URL \url{https://armadaproject.io/}.
\newblock [Online; accessed March 16, 2026].

\bibitem[Friese et~al.(2025)Friese, Eleliemy, Haus, and Schulz]{friese2025closinghpccloudconvergencegap}
Philipp~A. Friese, Ahmed Eleliemy, Utz-Uwe Haus, and Martin Schulz.
\newblock Closing the hpc-cloud convergence gap: Multi-tenant slingshot rdma for kubernetes, 2025.
\newblock URL \url{https://arxiv.org/abs/2508.09663}.

\bibitem[McIntosh-Smith et~al.(2024)McIntosh-Smith, Alam, and Woods]{mcintosh2024isambard}
Simon McIntosh-Smith, Sadaf Alam, and Christopher Woods.
\newblock Isambard-{AI}: a leadership-class supercomputer optimised specifically for artificial intelligence.
\newblock In \emph{Proceedings of the Cray User Group}, pages 44--54. ACM, 2024.

\bibitem[Watson(2025)]{watson2025code}
Jake Watson.
\newblock {FRIDGE} on {I}sambard-{AI}, 2025.
\newblock URL \url{https://github.com/isambard-sc/fridge}.

\bibitem[SATRE()]{satre}
SATRE.
\newblock {SATRE}, 2026.
\newblock URL \url{https://satre-specification.readthedocs.io/en/stable/}.

\bibitem[Moritz~\emph{et al}.(2018)]{moritz2018ray}
Philipp Moritz~\emph{et al}.
\newblock Ray: A distributed framework for emerging {AI} applications.
\newblock In \emph{13th USENIX Symposium on Operating Systems Design and Implementation (OSDI 18)}, pages 561--577. USENIX Association, 2018.

\bibitem[\emph{et al}.(2023)]{kwon2023}
Woosuk~Kwon \emph{et al}.
\newblock Efficient memory management for large language model serving with pagedattention.
\newblock In \emph{Proceedings of the ACM SIGOPS 29th Symposium on Operating Systems Principles}, 2023.

\bibitem[Kubernetes(2026{\natexlab{a}})]{dra}
Kubernetes.
\newblock Dynamic {R}esource {A}llocation, 2026{\natexlab{a}}.
\newblock URL \url{https://kubernetes.io/docs/concepts/scheduling-eviction/dynamic-resource-allocation/}.
\newblock [Online; accessed March 20, 2026].

\bibitem[Rancher(2026)]{rke2}
SUSE Rancher.
\newblock Rancher kubernetes engine 2, 2026.
\newblock URL \url{https://docs.rke2.io/}.
\newblock [Online; accessed March 30, 2026].

\bibitem[{T}uring {I}nstitute(2026)]{fridgearch}
Alan {T}uring {I}nstitute.
\newblock F{RIDGE}, 2026.
\newblock URL \url{https://github.com/alan-turing-institute/ fridge/blob/main/docs/architecture/architecture.md}.
\newblock [Online; accessed March 20, 2026].

\bibitem[DARE UK()]{dare}
DARE UK.
\newblock {DARE UK}, 2026.
\newblock URL \url{https://dareuk.org.uk/}.

\bibitem[FRIDGE()]{fridge}
FRIDGE.
\newblock {FRIDGE}, 2026.
\newblock URL \url{https://dareuk.org.uk/how-we-work/ongoing-activities/dare-uk-early-adopters/fridge/}.

\bibitem[Anyscale(2026)]{rayarch}
Anyscale.
\newblock Anyscale, 2026.
\newblock URL \url{https://www.anyscale.com/blog/ai-compute-open-source-stack-kubernetes-ray-pytorch-vllm}.
\newblock [Online; accessed March 23, 2026].

\bibitem[\emph{et al}.(2021)]{kanso2021}
Ali~Kanso \emph{et al}.
\newblock Designing a kubernetes operator for machine learning applications.
\newblock In \emph{Proceedings of the Seventh International Workshop on Container Technologies and Container Clouds}, 2021.
\newblock URL \url{https://api.semanticscholar.org/CorpusID:244663710}.

\bibitem[Wolf~\emph{et al}.(2019)]{wolf2019huggingface}
Thomas Wolf~\emph{et al}.
\newblock Huggingface's transformers: State-of-the-art natural language processing.
\newblock \emph{arXiv preprint arXiv:1910.03771}, 2019.
\newblock \doi{10.48550/arXiv.1910.03771}.
\newblock URL \url{https://arxiv.org/abs/1910.03771}.

\bibitem[Longhorn(2026)]{longhorn}
Longhorn.
\newblock {L}onghorn, 2026.
\newblock URL \url{https://longhorn.io/}.
\newblock [Online; accessed March 24, 2026].

\bibitem[NVIDIA(2026)]{nvidiagpuoperator}
NVIDIA.
\newblock {NVIDIA} {GPU} {O}perator, 2026.
\newblock URL \url{https://docs.nvidia.com/datacenter/cloud-native/gpu-operator/latest/overview.html/}.
\newblock [Online; accessed March 24, 2026].

\bibitem[{P}roject(2026{\natexlab{a}})]{kuberay}
Ray {P}roject.
\newblock Kube{R}ay {O}perator, 2026{\natexlab{a}}.
\newblock URL \url{https://github.com/ray-project/kuberay}.
\newblock [Online; accessed March 30, 2026].

\bibitem[Metacontroller(2026)]{metacontroller}
Metacontroller.
\newblock Metacontroller, 2026.
\newblock URL \url{https://github.com/metacontroller/metacontroller}.
\newblock [Online; accessed March 30, 2026].

\bibitem[SmarterDeviceManager()]{smarterdevicemanager}
SmarterDeviceManager.
\newblock {S}marter{D}evice{M}anager, 2026.
\newblock URL \url{https://github.com/smarter-project/smarter-device-manager}.

\bibitem[HPE CXI Kubernetes Device Plugin()]{hpek8sdeviceplugin}
HPE CXI Kubernetes Device Plugin.
\newblock {HPE CXI K}ubernetes {D}evice {P}lugin, 2026.
\newblock URL \url{https://github.com/HewlettPackard/cxi-k8s-device-plugin}.

\bibitem[Akri()]{akri}
Akri.
\newblock Akri, 2026.
\newblock URL \url{https://docs.akri.sh/}.

\bibitem[Pulumi(2026)]{pulumi}
Pulumi.
\newblock Pulumi {I}nfrastructure-as-{C}ode, 2026.
\newblock URL \url{https://www.pulumi.com/}.

\bibitem[MetalLB(2026)]{metallb}
MetalLB.
\newblock Metallb, 2026.
\newblock URL \url{https://metallb.io/}.
\newblock [Online; accessed March 23, 2026].

\bibitem[Cilium(2026)]{cilium}
Cilium.
\newblock Cilium, 2026.
\newblock URL \url{https://cilium.io/}.
\newblock [Online; accessed March 23, 2026].

\bibitem[Argo(2026)]{argoworkflows}
Argo.
\newblock Argo, 2026.
\newblock URL \url{https://argoproj.github.io/workflows/}.
\newblock [Online; accessed March 23, 2026].

\bibitem[{P}roject(2026{\natexlab{b}})]{raydocker}
Ray {P}roject.
\newblock Ray {Project} {D}ocker, 2026{\natexlab{b}}.
\newblock URL \url{https://hub.docker.com/layers/rayproject/ray/2.54.0-py312-cu126-aarch64/images/sha256-c64c52747e1540949f2340d1bee36c062a964123d7d601a3f803333e1bb2d059}.
\newblock [Online; accessed March 24, 2026].

\bibitem[Spack(2026)]{spack}
Spack.
\newblock Spack, 2026.
\newblock URL \url{https://spack.io/}.
\newblock [Online; accessed March 24, 2026].

\bibitem[Gamblin et~al.(2015)Gamblin, LeGendre, Collette, Lee, Moody, de~Supinski, and Futral]{10.1145/2807591.2807623}
Todd Gamblin, Matthew LeGendre, Michael~R. Collette, Gregory~L. Lee, Adam Moody, Bronis~R. de~Supinski, and Scott Futral.
\newblock The spack package manager: bringing order to hpc software chaos.
\newblock In \emph{Proceedings of the International Conference for High Performance Computing, Networking, Storage and Analysis}, SC '15, New York, NY, USA, 2015. Association for Computing Machinery.
\newblock ISBN 9781450337236.
\newblock \doi{10.1145/2807591.2807623}.
\newblock URL \url{https://doi.org/10.1145/2807591.2807623}.

\bibitem[Kubernetes(2026{\natexlab{b}})]{map}
Kubernetes.
\newblock Mutating {A}dmission {P}olicy, 2026{\natexlab{b}}.
\newblock URL \url{https://kubernetes.io/docs/reference/access-authn-authz/mutating-admission-policy/}.
\newblock [Online; accessed March 20, 2026].

\bibitem[OSU()]{osu}
OSU.
\newblock {OSU} {M}icro-{B}enchmarks, 7.5.
\newblock URL \url{https://mvapich.cse.ohio-state.edu/benchmarks/}.

\bibitem[Kyverno(2026)]{kyverno}
Kyverno.
\newblock Kyverno, 2026.
\newblock URL \url{https://kyverno.io/}.
\newblock [Online; accessed March 23, 2026].

\bibitem[KubeArmor(2026)]{kubearmor}
KubeArmor.
\newblock Kube{A}rmor, 2026.
\newblock URL \url{https://kubearmor.io/}.
\newblock [Online; accessed March 23, 2026].

\bibitem[{BentoML}(2026)]{Bento_ML:LLM_metrics}
{BentoML}.
\newblock Key metrics for {LLM} inference, 2026.
\newblock URL \url{https://bentoml.com/llm/inference-optimization/llm-inference-metrics}.
\newblock [Online; accessed March 12, 2026].

\bibitem[Cerebras(2024)]{cerebras2024llama405b}
Cerebras.
\newblock Llama 3.1 405b now runs at 969 tokens/s on cerebras inference, 2024.
\newblock URL \url{https://www.cerebras.ai/blog/llama-405b-inference}.

\bibitem[Oracle(2024)]{oracle2024mi300x}
Oracle.
\newblock Serving llama 3.1 405b model with amd instinct mi300x accelerators, 2024.
\newblock URL \url{https://blogs.oracle.com/cloud-infrastructure/serving-llama-31-405b-model-with-amd-mi300x-gpus}.

\bibitem[{vLLM}(2026)]{vLLM:benchmark_CLI}
{vLLM}.
\newblock v{LLM}: {B}enchmark {CLI}, 2026.
\newblock URL \url{https://docs.vllm.ai/en/latest/benchmarking/cli/}.
\newblock [Online; accessed March 13, 2026].

\bibitem[anon8231489123(2023)]{anon8231489123_sharegpt}
anon8231489123.
\newblock Sharegpt\_vicuna\_unfiltered, 2023.
\newblock URL \url{https://huggingface.co/datasets/anon8231489123/ShareGPT\_Vicuna\_unfiltered}.
\newblock Accessed: 2026-04-17.

\bibitem[{M}eta {L}lama(2026)]{metallama}
{M}eta {L}lama.
\newblock Meta-llama-3.1-405b-instruct, 2026.
\newblock URL \url{https://huggingface.co/meta-llama/Llama-3.1-405B-Instruct}.

\bibitem[BriCS(2026)]{brics_distributed_vllm_inference}
BriCS.
\newblock Distributed vllm inference, 2026.
\newblock URL \url{https://docs.isambard.ac.uk/user-documentation/tutorials/distributed-inference/}.
\newblock Accessed: 2026-04-20.

\bibitem[vLLM Authors(2026)]{vllm_distributed_multi_node_inference}
vLLM Authors.
\newblock vllm parallelism and scaling, 2026.
\newblock URL \url{https://docs.vllm.ai/en/stable/serving/parallelism_scaling/}.
\newblock Accessed: 2026-04-20.

\bibitem[USS()]{uss}
USS.
\newblock {USS -- User Services Software}, 2026.
\newblock URL \url{https://support.hpe.com/hpesc/public/docDisplay?docId=dp00005611en_us&page=install/About_USS.html&docLocale=en_US}.

\bibitem[{The Kubeflow Authors}(2026)]{kubeflow2026}
{The Kubeflow Authors}.
\newblock {Kubeflow}.
\newblock \url{https://www.kubeflow.org/}, 2026.
\newblock [Online; accessed 20-April-2026].

\bibitem[{The KServe Authors}(2026)]{kserve2026}
{The KServe Authors}.
\newblock {KServe}.
\newblock \url{https://kserve.github.io/website/}, 2026.
\newblock [Online; accessed 20-April-2026].

\end{thebibliography}
